\documentclass[runningheads]{llncs}

\usepackage[T1]{fontenc}
\usepackage{graphicx}
\usepackage{array}
\usepackage{float}
\usepackage{placeins}
\usepackage{svg}
\usepackage{url}
\usepackage{color}

\usepackage[
    colorlinks=true,
    urlcolor=black,
    linkcolor=black,
    citecolor=blue
]{hyperref}

\usepackage{orcidlink}

\begin{document}

\title{An Exploratory Pilot Survey on Technical Quality Control Practices in Agile R\&D Projects}
\titlerunning{Pilot Survey on Technical Quality in Agile R\&D}

\author{
Mateus Costa Lucena\orcidlink{0009-0001-2865-3930}
}

\authorrunning{Mateus C. Lucena}

\institute{
Venturus R\&D Institute, Manaus, Brazil \\
\email{mateus.lucena@venturus.org.br}
}

\maketitle

% ================================
% ABSTRACT
% ================================

\begin{abstract}
Managing technical quality in agile Research and Development (R\&D) software projects represents a persistent challenge, particularly in contexts characterized by high technical uncertainty and experimental pressure. This exploratory pilot survey explores how agile R\&D software teams report the use of practices and metrics related to technical quality control within Scrum-based environments. The study employed a structured questionnaire administered to professionals from Science and Technology Institutions (STIs) located in Manaus, Brazil, aiming to capture reported practices, perceptions of quality, and recurrent challenges. Quantitative data were complemented by qualitative responses to support contextual interpretation. The results indicate that although practices such as automated testing, code review, and continuous integration are widely acknowledged, their reported application is often inconsistent across iterations. Gaps were also observed in the monitoring of technical quality metrics and in the reporting of mechanisms for assessing technical debt from a business perspective. Rather than aiming for generalization, this study offers an exploratory baseline that describes how technical quality is managed in agile R\&D projects within a regional innovation ecosystem.
\end{abstract}


% ================================
% INTRODUCTION
% ================================

\section{Introduction}

Agile methodologies, consolidated since the publication of the Agile Manifesto~\cite{ref1}, have become one of the dominant paradigms for managing software development and Research and Development (R\&D) projects. Among these approaches, Scrum stands out due to its iterative structure, emphasis on transparency, and support for rapid feedback~\cite{ref2}. While agility enables faster adaptation to change, balancing delivery speed with sustained technical quality remains a persistent challenge~\cite{ref4}.

This tension is commonly associated with the accumulation of technical debt, defined by Cunningham~\cite{ref3} as the long-term cost incurred by short-term design or implementation decisions. When left unmanaged, technical debt increases maintenance effort, reduces system reliability, and constrains future system evolution~\cite{ref4,ref5}. In R\&D software projects, these risks are often amplified due to experimentation, technological novelty, and emergent requirements~\cite{ref16}. Traditional quality assurance approaches frequently prove too rigid for such contexts, while agile engineering practices are not always applied systematically or consistently~\cite{ref7,ref8,ref9,ref10}.

Recent empirical studies indicate that although agile practices such as automated testing, continuous integration, and code review are widely promoted, their consistent application remains uneven, particularly in innovation-driven environments~\cite{ref23}. Furthermore, despite the acknowledged importance of monitoring technical quality and technical debt, many teams struggle to operationalize metrics that translate technical concerns into business-relevant insights.

This study adopts an exploratory and descriptive approach to examine reported practices and perceptions related to technical quality control in agile R\&D software projects. Rather than prescribing normative solutions or attempting statistical generalization, the study focuses on capturing self-reported practices, perceived challenges, and contextual constraints experienced by practitioners. The research is positioned as a pilot investigation intended to provide an empirical baseline and to inform the design of more comprehensive future studies.

The research was conducted within the Polo Industrial de Manaus (PIM), a major Brazilian innovation hub supported by the Brazilian Informatics Law (No.~8.248/1991), which mandates private investment in R\&D activities~\cite{EldoradoLeiInformatica}. While the study is situated in an R\&D context, the analysis focuses on reported practices and perceptions rather than on measuring R\&D-specific process characteristics. This setting nevertheless provides a relevant context for examining how agile teams approach technical quality under conditions of experimentation, regulatory incentives, and delivery pressure.

The following research questions are exploratory in nature and focus on self-reported practices and perceptions:

\begin{itemize}
    \item (RQ1) How do agile R\&D software teams report the application of technical quality control practices within Scrum-based projects?
    \item (RQ2) Which technical quality metrics are reported as being used, and how consistently are they monitored?
    \item (RQ3) What challenges and barriers are reported as affecting the implementation of technical quality practices in agile R\&D environments?
    \item (RQ4) How are scope changes reported to influence the perception and impact of technical debt during project execution?
\end{itemize}

This paper is structured as follows. Section~2 presents related work on quality management, agile methods, and technical debt. Section~3 details the research methodology. Section~4 reports and discusses the empirical findings. Section~5 concludes the paper and outlines directions for future research.

To ensure analytical transparency, RQ1 and RQ2 are primarily addressed through quantitative survey data (Sections~4.2 and~4.3), while RQ3 and RQ4 are explored using both quantitative and qualitative responses (Sections~4.4 and~4.5).

% ================================
% RELATED WORK (FULL – PRESERVED)
% ================================

\section{Related Work}

In R\&D environments, quality principles acquire heightened importance due to the experimental nature of activities and the technological uncertainty inherent to such projects~\cite{isoneworld2017}. Prior studies indicate that the success of R\&D initiatives depends not only on their capacity for innovation but also on the reliability, traceability, and technical sustainability of their deliverables~\cite{ref15}. In this context, quality control mechanisms must remain iterative and adaptable in order to reconcile experimentation with a minimum level of predictability~\cite{ref16}.

The literature suggests that technical quality in innovation-driven environments is often managed reactively rather than preventively, particularly when rapid prototyping is prioritized over systematic monitoring and documentation~\cite{mappingTD}. This reactive orientation limits teams’ ability to anticipate the long-term effects of architectural and implementation decisions, increasing the risk of technical debt accumulation~\cite{ref4}. Such dynamics are frequently reported in contexts where short-term delivery pressures coexist with high technical uncertainty.

Since the publication of the Agile Manifesto~\cite{ref1}, frameworks such as Scrum~\cite{ref2} have been widely adopted to support short delivery cycles, continuous feedback, and rapid adaptation to change. Empirical studies, however, point to a recurring tension: although agile methods facilitate responsiveness, technical quality may degrade when engineering practices and monitoring mechanisms lack consistency across iterations~\cite{ref7,scrumQuality2022,refCartaxo2013}. Evidence suggests that Scrum teams achieve better quality outcomes when practices such as automated testing and structured code reviews are applied in a disciplined and sustained manner~\cite{scrumQuality2022}.

Scaled agile frameworks, including SAFe~\cite{ref10} and LeSS~\cite{ref22}, were introduced to address coordination and governance challenges in multi-team environments. While these frameworks primarily aim to support agility at scale, they also introduce structures intended to reinforce traceability and quality assurance~\cite{ref9}. Empirical findings indicate, however, that the effectiveness of such mechanisms strongly depends on organizational factors such as leadership alignment, team autonomy, and technical maturity~\cite{ref23,ref24}, reinforcing the context-dependent nature of quality management practices.

Prior work also identifies multiple categories of metrics used to monitor quality in agile projects. Process metrics, such as velocity and lead time, support workflow evaluation~\cite{ref7}. Technical metrics, including test coverage, static analysis findings, and defect density, are commonly used to assess structural quality attributes~\cite{ref4,ref18}. Product-level metrics, such as availability and reliability, provide insight into system behavior in operation~\cite{ref5,ref23}. Despite the availability of these indicators, studies consistently report difficulties in sustaining continuous measurement, particularly in R\&D environments characterized by experimentation and evolving requirements~\cite{mappingTD,ref17}.

Recent empirical research further highlights that visualization techniques for technical debt management remain underutilized, even though they are widely perceived as beneficial. Irgens and Martini~\cite{IrgensMartini2025} report that only a limited number of teams systematically employ visualization tools to support technical debt analysis and prioritization, revealing a persistent gap between perceived value and practical adoption.

Overall, the literature converges on two central observations that motivate this study. First, technical quality in agile R\&D projects is broadly acknowledged as essential but is inconsistently monitored in practice. Second, the sustainability of quality practices depends heavily on organizational context and governance structures. These observations reinforce the relevance of exploratory empirical studies that examine how technical quality practices are reported and perceived in specific innovation ecosystems.

% ================================
% RESEARCH METHODOLOGY (FULL)
% ================================

\section{Research Methodology}

This study followed an exploratory pilot mixed-methods approach, combining quantitative and qualitative data to capture self-reported practices and perceptions related to technical quality control in agile R\&D software projects. 

\begin{figure}
\centering
\includegraphics[width=0.88\textwidth]{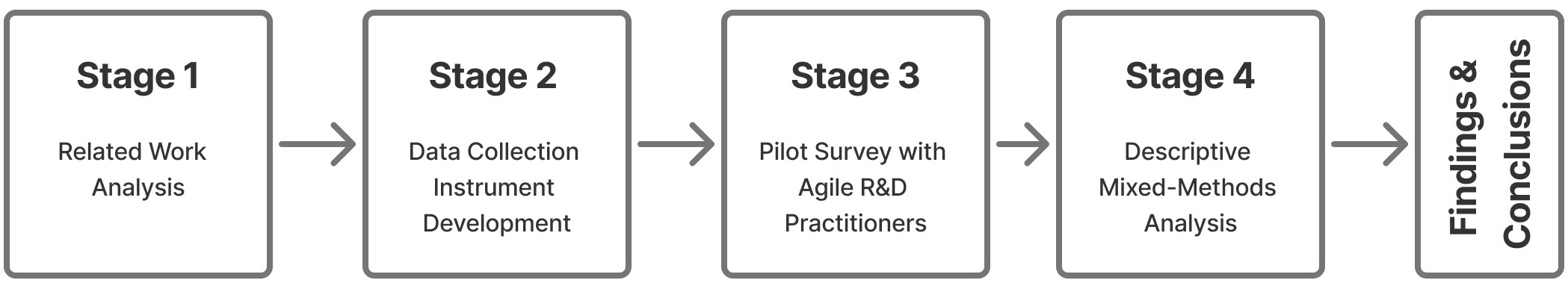}
\caption{Research design and methodological stages of the exploratory pilot study.}
\label{fig:methodology}
\end{figure}

Figure~\ref{fig:methodology} summarizes the main methodological stages. The process began with a focused review of the literature (Section~2), which informed the definition of the survey topics and the formulation of the questionnaire items. A structured questionnaire was then administered to practitioners to collect empirical data.

The survey instrument was intentionally designed to prioritize breadth over depth. Rather than providing a fine-grained operationalization of specific practices or metrics, the questionnaire aimed to obtain an initial overview of reported practices, perceived challenges, and gaps in technical quality management. This design choice is consistent with the exploratory nature of the study and with its role as an empirical baseline intended to inform subsequent, more detailed investigations.

\newpage

\subsection{Sample and Participants}

The sample consisted of professionals from Science and Technology Institutions (STIs) involved in software-related activities within the Manaus-AM region. Participants reported working in projects associated with R\&D initiatives and represented diverse roles, including software developers, project leaders, project managers, quality professionals, and business analysts.

The questionnaire was distributed in January 2025 to 40 professionals who had previously attended an agile leadership workshop conducted by the author. Participation was voluntary and based on convenience sampling. A total of 17 valid responses were obtained. Agile experience and involvement in R\&D activities were inferred from participants’ organizational context and self-reported project experience, rather than being explicitly measured through dedicated survey items. As such, the study relies on participants’ perceptions and reported experiences.

All responses were collected anonymously. To support transparency and replicability, the complete raw dataset is publicly available for inspection at:
\href{https://docs.google.com/spreadsheets/d/e/2PACX-1vQVcHJsMSsGdgVWCwArr62AdLxRLkpkKgLS31QVUYhk6efQQ6FMhzO0i0mbf3PhOp4wOrv4fuRv5Afa/pubhtml?gid=692209400&single=true}{Survey Results}.

\subsection{Data Collection}

Data were collected using an online questionnaire comprising 11 questions. The questionnaire included both closed-ended and open-ended items and was organized into the following thematic domains:

\begin{itemize}
    \item Areas of Activity
    \item Quality Control Tools and Integration
    \item Technical Quality Metrics
    \item Challenges and Barriers
    \item Impact of Technical Debt
\end{itemize}

Closed-ended questions were used to capture frequencies and distributions of reported practices, while open-ended questions allowed participants to elaborate on perceived challenges and the impact of technical debt in their project contexts.

The full structure of the questionnaire, including all questions and predefined response options, is presented in Table~\ref{tab2}. Multiple selections were allowed in questions related to roles, project domains, and quality practices, as indicated in the results section.

\newpage

% === TABLE (UNCHANGED) ===
% [TABLE CONTENT REMAINS EXACTLY AS IN YOUR ORIGINAL FILE]
\renewcommand{\arraystretch}{2.9}
\begin{table}[H]
\caption{Technical Quality Assessment Form.}
\label{tab2}
\centering
\resizebox{\textwidth}{!}{
\begin{tabular}{>{\raggedright\arraybackslash}p{3.8cm}
                >{\raggedright\arraybackslash}p{7.2cm}
                >{\raggedright\arraybackslash}p{5.7cm}}
\hline
\textbf{Section} & \textbf{Questions} & \textbf{Alternatives} \\
\hline

\textbf{Section 1: Areas of Activity} &
1. What is your role within the agile development team? &
(a) Developer \newline
(b) Quality Professional \newline
(c) Project Leader \newline
(d) Business Analyst \newline
(e) Project Manager \newline
(f) Other \\

&
2. What is the general theme of the projects you have participated in? &
(a) Manufacturing Systems \newline
(b) Industry 4.0 \newline
(c) Communication Systems \newline
(d) Payment Systems \newline
(e) Data Systems \newline
(f) AI Systems \newline
(g) Other \\

\textbf{Section 2: Quality Control Tools and Integration} &
3. Which quality control tools or methods were used during project development? &
(a) Automated Testing \newline
(b) Code Review \newline
(c) Integration Testing \newline
(d) None \newline
(e) Other \\

&
4. How often were quality control tools or methods integrated into agile iterations in your projects? &
(a) In every iteration \newline
(b) Frequently, but not in all \newline
(c) Rarely \newline
(d) Never \\

\textbf{Section 3: Technical Quality Metrics} &
5. In the projects you participated in, were technical metrics monitored to ensure product quality across all knowledge areas? &
(a) Yes \newline
(b) No \newline
(c) Other \\

\textbf{Section 4: Challenges and Barriers} &
6. How often did your team rely on external resources to balance quality and agility during development? &
(a) In every iteration \newline
(b) Frequently, but not in all \newline
(c) Rarely \newline
(d) Never \\

&
7. What do you consider to be the main challenge faced by teams when applying practices that ensure technical quality in agile development environments? &
(a) Lack of Process Standardization \newline
(b) Frequent Scope Changes \newline
(c) Time Limitations for Experimentation/Testing \newline
(d) Lack of Resources or Managerial Support \\

&
8. To what extent do you agree with the statement: “The Scrum methodology allows maintaining high quality standards without compromising delivery speed”? &
(a) Strongly Agree \newline
(b) Partially Agree \newline
(c) Partially Disagree \newline
(d) Strongly Disagree \\

\textbf{Section 5: Impact of Technical Debt} &
9. How do you assess the impact of scope changes on the implementation of technical quality assurance practices in agile projects? &
Open-ended question \\

&
10. In the projects you participated in, were there any indicators used to measure the impact of technical debt on deliveries from a business perspective? &
Open-ended question \\

&
11. General Comments (Optional) &
Open-ended question \\
\hline
\end{tabular}
}
\end{table}

% ================================
% RESULTS AND DISCUSSION (FULL, ALL FIGURES PRESERVED)
% ================================

\section{Results and Discussion}

Given its exploratory pilot nature, the survey results are interpreted as descriptive evidence of self-reported practices rather than as statistically generalizable findings. Results are presented following the structure of the questionnaire, and interpretation is explicitly grounded in reported frequencies and qualitative statements. Where applicable, findings are discussed in light of existing literature to contextualize observed patterns.

\subsection{Areas of Activity}

This subsection provides contextual information to support RQ1 by characterizing the professional roles and project domains reported by participants.

\begin{figure}[H]
\centering
\includegraphics[width=0.67\textwidth]{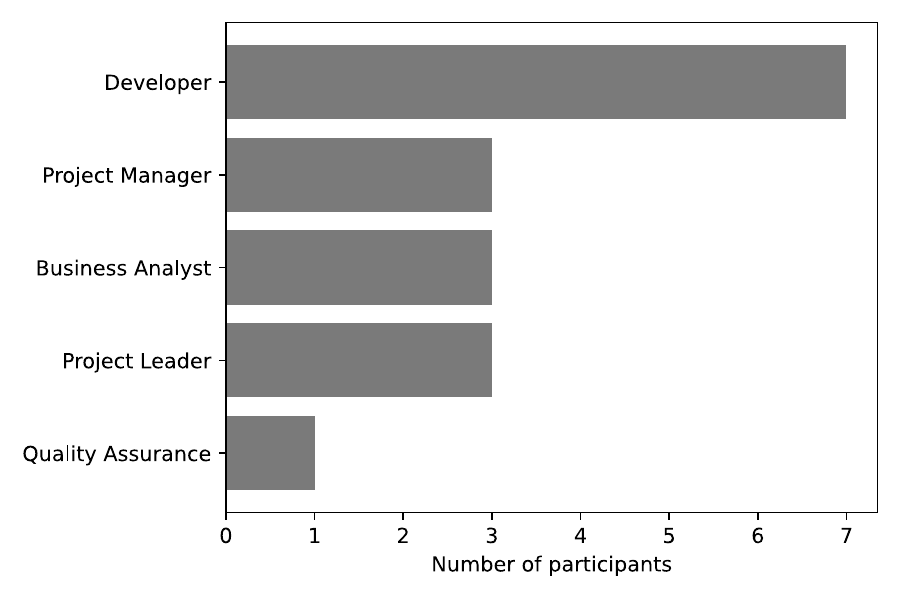}
\caption{Distribution of participants roles in agile development teams (N = 17).}
\label{fig:roles}
\end{figure}

Figure~\ref{fig:roles} presents the distribution of participant roles within agile teams. Among the 17 respondents, the majority reported acting primarily as Developers, followed by Project Leads, Project Managers, and Business Analysts. No single role accounted for an absolute majority of responses, indicating some diversity in professional perspectives within the sample. However, quality-focused roles were less represented, suggesting a potential bias toward development-centric viewpoints and highlighting an opportunity for broader role inclusion in future studies.

\begin{figure}
\centering
\includegraphics[width=0.67\textwidth]{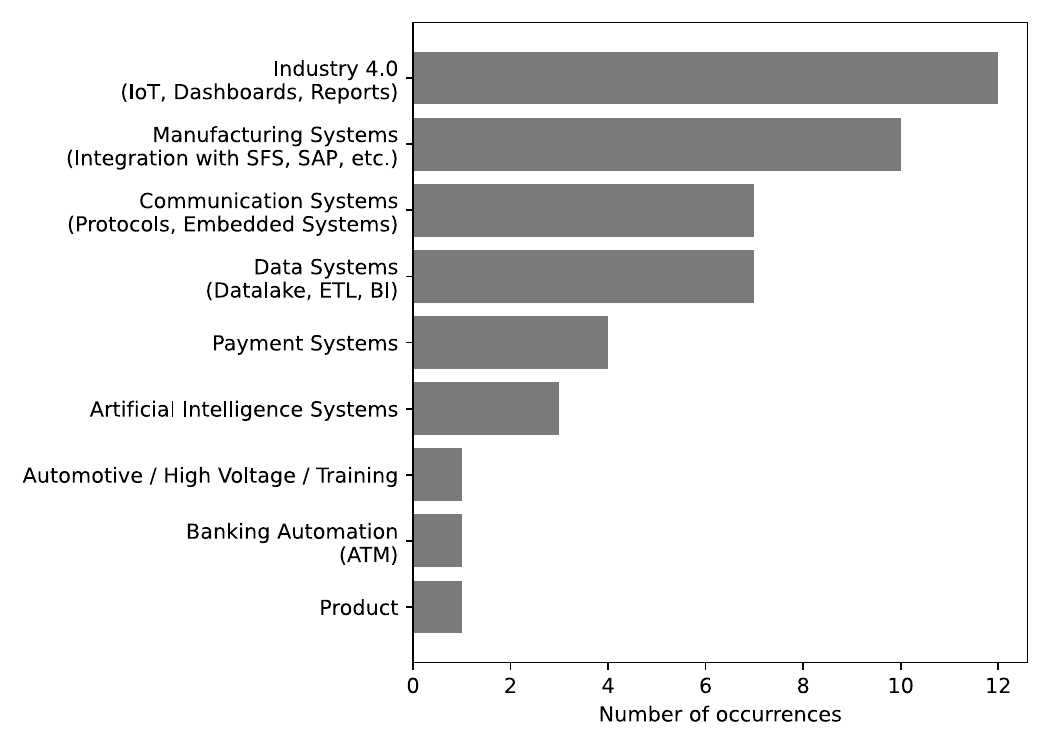}
\caption{Distribution of project themes by part.(multiple selections allowed, N = 17).}
\label{fig:themes}
\end{figure}

Figure~\ref{fig:themes} summarizes the project domains reported by participants. Industry~4.0 and Digital Manufacturing were the most frequently selected domains, followed by Communication Systems, Digital Payment Systems, Big Data, and Artificial Intelligence. Multiple selections were allowed. This distribution reflects the technological diversity typically associated with R\&D-oriented environments in the analyzed ecosystem.

\subsection{Quality Control Tools and Integration}

This subsection addresses RQ1 by describing the technical quality control practices reported as being used within Scrum-based agile projects.

\begin{figure}[H]
\centering
\includegraphics[width=0.67\textwidth]{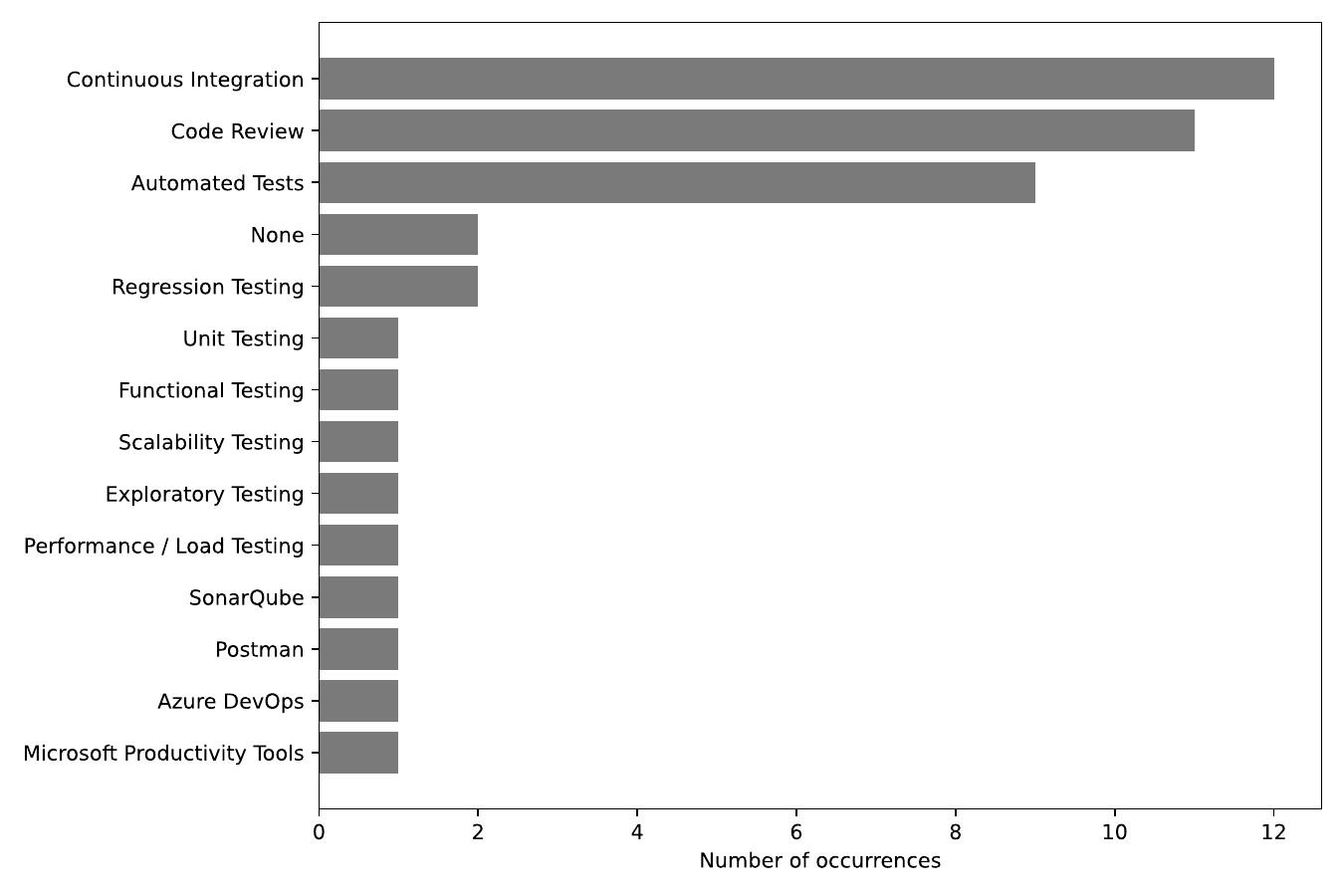}
\caption{Distribution quality control tools and methods (multiple selections allowed).}
\label{fig:tools}
\end{figure}

Figure~\ref{fig:tools} shows the distribution of quality control tools and methods reported by participants. Automated testing and code review were the most frequently cited practices, followed by continuous integration. These practices are commonly associated with early defect detection and shorter feedback loops in agile development, as reported in prior empirical studies~\cite{ref16,ref4}.

Other practices, including QA testing, regression testing, and the use of tools such as SonarQube, were reported by one or two participants each (approximately 5.9\%). This suggests that these practices are applied selectively in specific projects rather than being systematically adopted across teams.

\begin{figure}[H]
\centering
\includegraphics[width=0.67\textwidth]{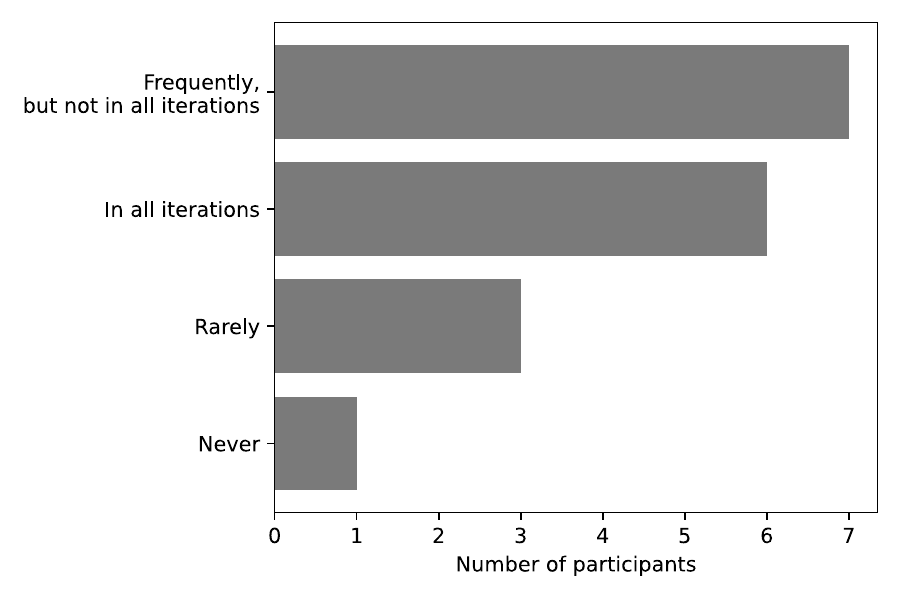}
\caption{Reported frequency of integration of quality control tools in agile iterations.}
\label{fig:frequency}
\end{figure}

Figure~\ref{fig:frequency} presents how frequently quality control tools were reported as being integrated into agile iterations. Seven out of 17 participants (41.2\%) indicated that such tools were used frequently but not in every iteration. Five participants (29.4\%) reported usage in every iteration, while one participant (5.9\%) reported that quality control tools were never used. These results indicate that, even when practices are acknowledged, their application tends to be situational rather than continuous.

\subsection{Technical Quality Metrics}

This subsection addresses RQ2 by examining which technical quality metrics are reported as being used and how consistently they are monitored across projects.

\begin{figure}[H]
\centering
\includegraphics[width=0.67\textwidth]{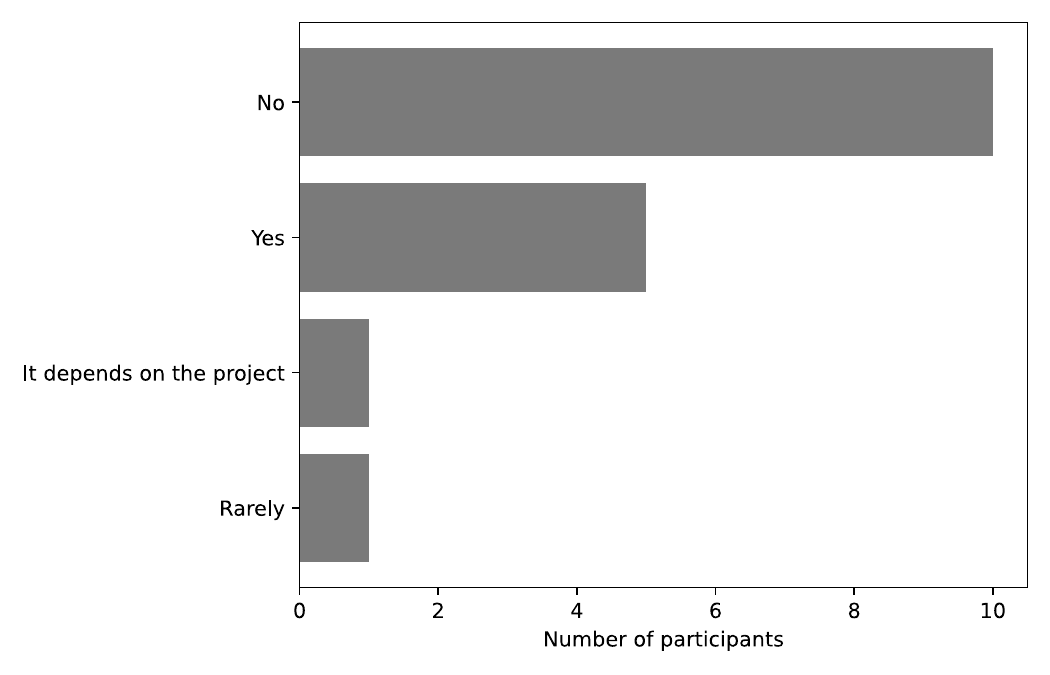}
\caption{Self-reported monitoring of technical quality metrics across areas (N = 17).}
\label{fig:metrics}
\end{figure}

Figure~\ref{fig:metrics} summarizes responses related to the monitoring of technical quality metrics. The majority of participants reported that technical metrics were not monitored consistently across all relevant project areas. This finding points to limited formalization and continuity of metric-based monitoring, particularly in projects involving multiple technical domains.

Although some form of metric monitoring was reported, participants rarely mentioned mechanisms for linking technical indicators to business-level outcomes. This gap limits the use of metrics as decision-support tools for prioritizing technical improvements.

\newpage

\begin{figure}[H]
\centering
\includegraphics[width=0.67\textwidth]{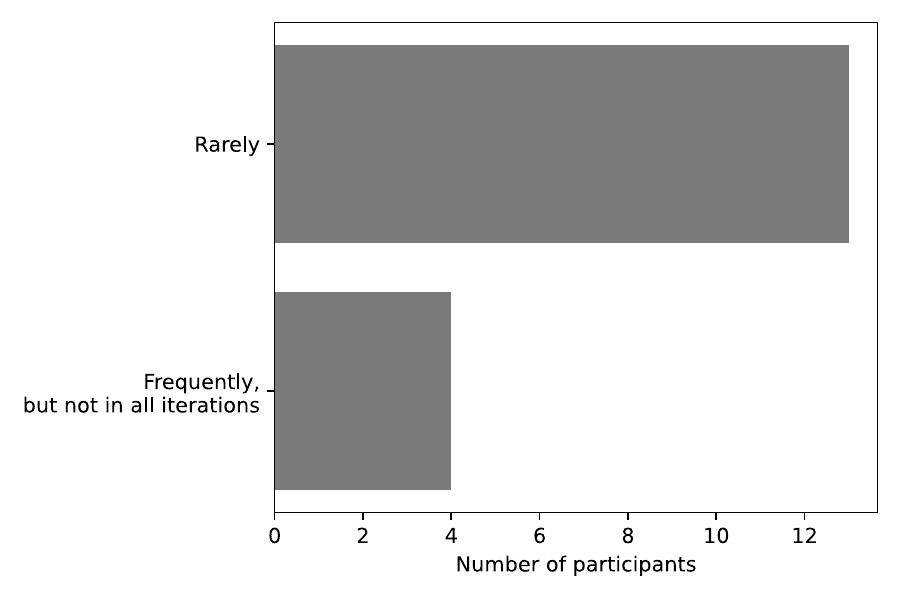}
\caption{Self-reported use of external quality support during development (N = 17).}\label{fig:external}
\end{figure}

Figure~\ref{fig:external} shows how frequently teams reported relying on external resources, such as architects or consultants, to support quality-related activities. Eight out of 17 participants (47.1\%) reported frequent or continuous reliance on external expertise. This pattern is consistent with observations in large-scale or technically complex agile environments, where specialized knowledge is required to address architectural concerns and accumulated technical debt~\cite{ref10}.

\newpage

\subsection{Challenges and Barriers}

This subsection addresses RQ3 by presenting the main challenges and barriers reported as affecting the implementation of technical quality practices in agile R\&D environments.

\begin{figure}[H]
\centering
    \includegraphics[width=0.64\textwidth]{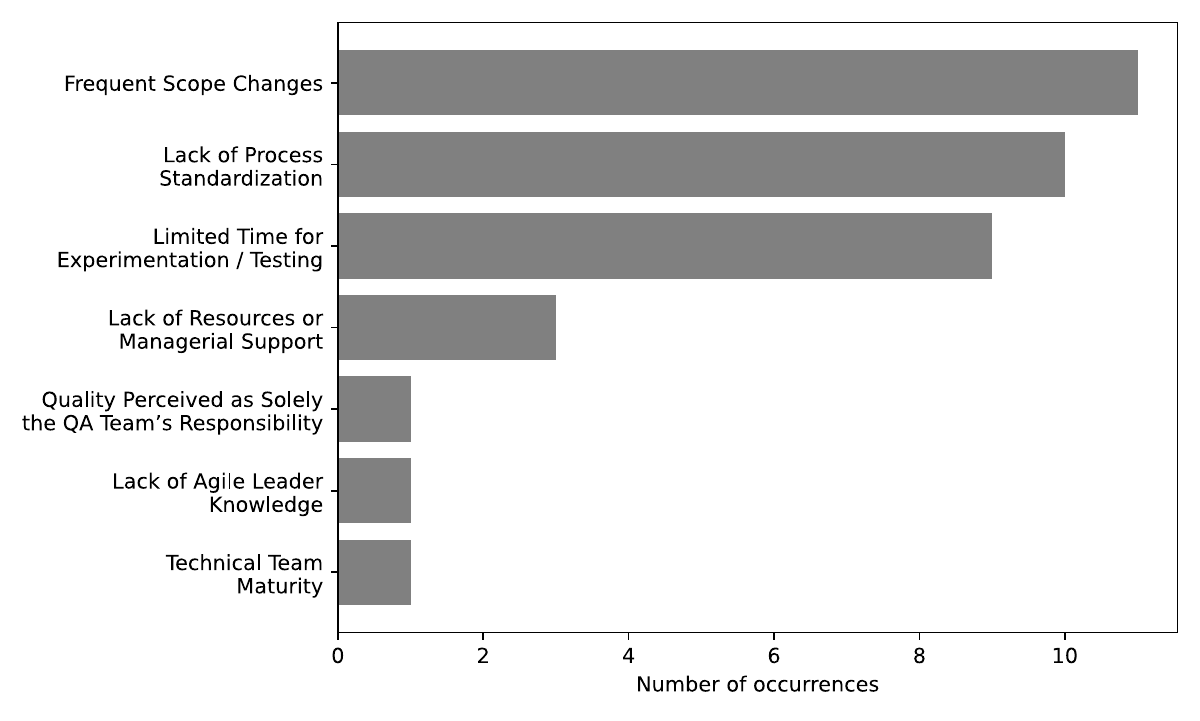}
\caption{Reported challenges in agile R\&D product development.}
\label{fig:challenges}
\end{figure}

Figure~\ref{fig:challenges} presents the main challenges reported by participants. Frequent scope changes were identified as the most common barrier affecting technical quality. While adaptability is a core principle of agile development, repeated scope changes were reported to increase rework and disrupt planned testing and refactoring activities.

\begin{figure}[H]
\centering
\includegraphics[width=0.54\textwidth]{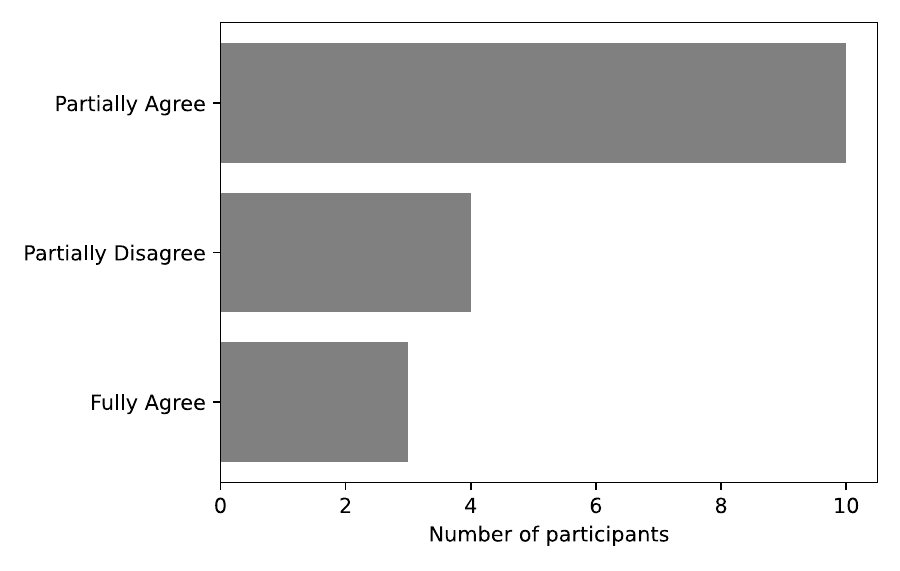}
\caption{Perceived effectiveness of Scrum in keep quality without compromising agility.}
\label{fig:scrum}
\end{figure}

Figure~\ref{fig:scrum} summarizes participants’ perceptions regarding Scrum’s ability to maintain quality without compromising delivery speed. Twelve out of 17 participants (70.6\%) reported agreement with this statement, while four participants (23.5\%) partially disagreed. No participant strongly disagreed. This suggests a generally positive perception of Scrum, accompanied by recognition of practical limitations in its application.

\subsection{Impact of Technical Debt}

This subsection addresses RQ4 by analyzing how scope changes are reported to influence the perception and impact of technical debt during project execution. Open-ended responses were analyzed using a lightweight thematic grouping, resulting in three recurring topics.

\subsubsection{Impact of Scope Changes}

Ten out of 17 participants explicitly reported that scope changes negatively affect project cost and schedule. Respondents indicated that late changes often require additional testing or rework, reducing the time available for preventive quality practices. While replanning is viewed as an inherent aspect of agile development, excessive scope volatility was associated with increased technical debt and reduced efficiency.

\subsubsection{Absence of Indicators for Measuring Technical Debt}

Twelve participants reported the absence of formal indicators for assessing the business impact of technical debt. Instead, technical debt was described as being perceived through informal discussions or subjective judgment. A small number of respondents mentioned indirect indicators, such as reduced velocity or increased rework, but these were not associated with systematic mitigation strategies.

\subsubsection{Challenges in Sustaining Quality Practices}

Participants reported organizational and technical barriers to sustaining quality practices, including limited process standardization and difficulties in maintaining consistent practices across iterations. Communication gaps and misaligned expectations among stakeholders were also cited. The notion of \textit{fitness-for-use}~\cite{ref23} was mentioned as a pragmatic way of negotiating quality expectations under constrained resources. Consistent with prior studies~\cite{IrgensMartini2025}, participants did not report a lack of perceived value in visualization tools, but rather difficulties in embedding technical indicators into routine decision-making processes.

% ================================
% THREATS TO VALIDITY (NEW)
% ================================

\section{Threats to Validity}

As an exploratory pilot study, this research is subject to limitations that should be considered when interpreting the results. Threats to validity are discussed following commonly adopted categories in empirical software engineering research.

\textbf{Construct validity} is primarily affected by the reliance on self-reported data. The survey captures participants’ perceptions and recollections of technical quality practices rather than direct observations of development artifacts or objectively measured behaviors. Core concepts such as technical quality, technical debt, and consistency of practice adoption may have been interpreted differently by respondents. In addition, agile experience and involvement in R\&D activities were inferred from organizational context and self-reported project experience rather than explicitly measured, which may introduce variability in respondents’ frames of reference.

\textbf{Internal validity} may be influenced by recall and interpretation bias. Participants may emphasize recent or particularly salient project experiences while underrepresenting long-term or less visible quality issues. Due to the cross-sectional nature of the survey, the study does not support causal claims regarding relationships between scope changes, quality practices, and technical debt. Accordingly, the analysis focuses on reported associations and perceived impacts rather than causal explanations.

\textbf{External validity} is limited by the small sample size (N = 17) and the concentration of participants within a single regional innovation ecosystem. Although the Polo Industrial de Manaus represents an active R\&D context, the findings are context-dependent and should not be generalized to other organizational, industrial, or geographic settings. The use of convenience sampling further constrains generalizability.

\textbf{Conclusion validity} is supported by the descriptive and exploratory stance adopted throughout the study. Quantitative results are reported using descriptive statistics only, and qualitative responses are used to complement and contextualize these findings. While consistency between quantitative trends and qualitative themes provides internal coherence, the results should be interpreted as indicative patterns rather than definitive conclusions.

% ================================
% CONCLUSIONS (FULL)
% ================================

\section{Conclusions and Future Work}

This exploratory pilot survey examined how professionals from STIs in Manaus, Brazil, report the use of technical quality control practices in agile R\&D software projects, with a particular focus on Scrum-based environments. The results indicate that practices such as automated testing, code review, and continuous integration are widely acknowledged by participants. However, their reported application is often inconsistent across projects and iterations, suggesting challenges in sustaining technical quality practices in experimentation-driven contexts.

The findings also point to limited use of standardized indicators for monitoring technical quality and technical debt. Participants rarely reported mechanisms that explicitly connect technical indicators to business-level outcomes, which restricts the use of metrics as systematic decision-support tools. Frequent scope changes, time constraints, and reliance on specialized external resources were reported as factors contributing to fragmented quality practices, reinforcing the context-dependent nature of quality management in agile R\&D environments.

Consistent with prior work, the results suggest that visualization techniques for technical debt are perceived as valuable but are not easily embedded into routine decision-making processes. Rather than indicating a lack of tools, this pattern highlights difficulties in integrating technical quality discussions into existing agile ceremonies and governance structures.

Several limitations must be considered when interpreting these findings. The study relied on a small, convenience-based sample drawn from a single regional innovation ecosystem, and all data were self-reported. As a result, the findings should be interpreted as indicative patterns rather than generalizable evidence. Nevertheless, the diversity of professional roles represented supports the relevance of the observations as exploratory input.

Future research should build on this pilot investigation through in-depth case studies and interviews with multidisciplinary teams to better understand how technical quality practices evolve throughout agile R\&D project lifecycles. Additional studies may also refine and validate survey instruments by explicitly measuring participant experience, distinguishing between process-, technical-, and product-level metrics, and clarifying the intended decision-making purposes of these indicators. Overall, this study provides an initial empirical baseline to support further investigation into technical quality management in agile R\&D contexts.

\subsubsection*{Acknowledgments}

The author thanks Venturus R\&D Institute for supporting applied research in agile software development. He also thanks Professor Antonio Carlos Pacagnella Junior for his valuable feedback and encouragement provided after the initial pilot study, which helped refine the research positioning and future directions.

\bibliographystyle{splncs04}
\bibliography{references}

\end{document}